\documentclass[sigconf,screen]{acmart}

\AtBeginDocument{%
  \providecommand\BibTeX{{%
    \normalfont B\kern-0.5em{\scshape i\kern-0.25em b}%
    \kern-0.8em\TeX%
  }}%
}

\copyrightyear{2026}
\acmYear{2026}
\setcopyright{cc}
\setcctype{by}

\acmConference[NordiCHI '26 Adjunct]
{Adjunct Proceedings of the 14th Nordic Conference on Human-Computer Interaction}
{October 3--7, 2026}
{Vaasa, Finland}

\acmBooktitle{Adjunct Proceedings of the 14th Nordic Conference on
Human-Computer Interaction (NordiCHI '26 Adjunct),
October 3--7, 2026, Vaasa, Finland}

\acmDOI{10.1145/3821402.3830117}
\acmISBN{979-8-4007-2798-6/2026/10}

\usepackage{booktabs}
\usepackage{dcolumn}
\usepackage{multirow}
\usepackage{makecell}
\usepackage{array}

\usepackage{framed}
\usepackage{enumitem}
\usepackage{lscape}
\usepackage{subcaption}
\usepackage{tikz}

\usepackage{etoolbox}

\newcolumntype{M}[1]{%
  >{\centering\arraybackslash}m{#1}%
}

\newcolumntype{L}[1]{%
  >{\raggedright\arraybackslash}p{#1}%
}

\newcolumntype{C}[1]{%
  >{\centering\arraybackslash}p{#1}%
}

\begin{document}


\title[Digital Tensions in Older Adults' Use of Norwegian Web Services]
{Legally Mandated, but Still Inaccessible: Digital Tensions in Older
Adults' Use of Norwegian Web Services}

\author{Yavuz Inal}
\authornote{Corresponding author.}
\email{yavuz.inal@ntnu.no}
\orcid{0000-0001-9919-6637}
\affiliation{%
  \department{Department of Design}
  \institution{Norwegian University of Science and Technology}
  \city{Gjøvik}
  \country{Norway}
}

\author{Sujay Shalawadi}
\email{sujay.shalawadi@ntnu.no}
\orcid{0000-0003-3937-5427}
\affiliation{%
  \department{Department of Design}
  \institution{Norwegian University of Science and Technology}
  \city{Gjøvik}
  \country{Norway}
}

\author{Eleftherios Papachristos}
\email{eleftherios.papachristos@ntnu.no}
\orcid{0000-0003-2666-9864}
\affiliation{%
  \department{Department of Design}
  \institution{Norwegian University of Science and Technology}
  \city{Gjøvik}
  \country{Norway}
}

\renewcommand{\shortauthors}{Inal, Shalawadi, and Papachristos}


\begin{abstract}
Norway is among the most digitalized countries in the world, where
access to essential services increasingly depends on digital systems.
Although universal design of ICT is legally required across public and
private sectors, ensuring cognitive accessibility for older adults
involves more than technical compliance. We analyzed responses from
294 participants aged 55 to 90 to examine the barriers they encounter
when using digital services. Our findings identify four tensions,
namely navigational, semantic, procedural, and temporal, which reveal
how accessibility barriers emerge through misalignments between service
design, system assumptions, and older adults' capabilities. Together,
these tensions show that checklist-based compliance does not necessarily
translate into lived accessibility when engaging with mandatory or
near-mandatory digital services. We discuss the need to move beyond
minimum compliance frameworks toward accessibility approaches that
better support older adults' independent participation in digital
society.
\end{abstract}




\maketitle


\section{Introduction}

Digital services have become a primary gateway to everyday participation in Norway and other Scandinavian countries. Banking, healthcare, taxation, welfare, shopping, and municipal services are increasingly accessed through websites and apps rather than physical or human-mediated channels. In Norway, universal design of Information and Communication Technology (ICT) is legally required for both public and private sector services, and websites are expected to comply with Web Content Accessibility Guidelines (WCAG) level AA requirements~\cite{uutilsynet2025wcag}. For older adults, however, accessibility is not only about meeting technical requirements. Ageing can involve changes in vision, dexterity, memory, attention, confidence, and recovery from errors, all of which shape whether a service can actually be used in practice~\cite{w3c2025olderusers}. Older adults' accessibility problems are often related to comprehension, navigation, and task completion rather than isolated interface defects alone~\cite{sayago2011ethnographical}.

This generates a gap between formal accessibility compliance and accessible use in practice. WCAG level AA compliance is mandatory and intended to ensure accessible digital services, yet compliance does not necessarily guarantee accessible use. WCAG provides an essential baseline, but many barriers experienced by older adults are difficult to capture through automated checks or checklist-based conformance alone~\cite{petrie2007relationship}. Although WCAG 2.2 addresses visual, physical, and cognitive accessibility~\cite{w3c2024wcag22}, many issues relevant to older adults cannot be fully assessed through automated conformance testing alone and often require interpretation and user testing in practice ~\cite{sayago2011ethnographical,petrie2007relationship}. As a result, websites may achieve high accessibility scores while still remaining difficult, stressful, or exclusionary for older adults. Our findings suggest that mandatory AA-level compliance is necessary, but insufficient for ensuring accessible use in practice.

In this paper, we examine this tension through an exploratory survey of older adults’ experiences with Norwegian web services that are legally expected to be accessible. We collected responses from 294 participants aged 55 to 90 regarding the services they use, barriers they encounter, and factors preventing more frequent use. We also evaluated representative Norwegian websites across different service levels. These websites achieved high automated accessibility scores, with mean scores of 97.9 for high-usage services, 97.4 for moderate-usage services, and 92.7 for low-usage services. Despite this, participants reported barriers related to privacy and security concerns, service complexity, lack of support, fear of mistakes, visual difficulties, and health-related challenges.

This paper makes three contributions. First, it shows that high AA-oriented accessibility scores do not necessarily reflect older adults’ lived accessibility when using mandatory or near-mandatory digital services. Second, it identifies four forms of digital tension that explain how barriers emerge across the user journey: navigational, semantic, procedural, and temporal. Third, it provides empirically grounded insights that can inform designers, developers, service owners, and policymakers seeking to move beyond minimum compliance toward services older adults can actually understand, trust, and use independently.

\section{Related Work}

\subsection{Accessibility Compliance and the Limits of Guideline-Based Evaluation}

Web accessibility is commonly framed in terms of technical standards and regulatory mechanisms, such as the Web Content Accessibility Guidelines (WCAG). These frameworks have become central to accessibility audits and automated evaluation practices, particularly in public-facing digital services. However, prior work shows that compliance with accessibility standards does not necessarily ensure meaningful accessibility in practice~\cite{Inal2022}. Designers and developers often struggle to translate accessibility principles into concrete design and development decisions due to limited awareness and competing priorities~\cite{antonelli2018survey, farrelly2011practitioner}. Similarly, UX professionals recognize the importance of accessibility but report limited familiarity with accessibility standards and spend relatively little time addressing accessibility during projects~\cite{inal2019web, Inal}. As a result, accessibility concerns may remain overlooked throughout the development lifecycle, even when guidelines and evaluation tools are available.

Similar challenges have also been identified in accessibility education research, where students often struggle to move beyond applying guidelines toward understanding how accessibility issues are experienced in practice~\cite{katsanos2009web, inal2018computer, soares2020we}. Consequently, accessibility education has increasingly emphasized experiential and empathy-based approaches that expose students to barriers encountered by individuals with disabilities~\cite{putnam2016best, ludi2018teaching, el2020presenting}. Yet opportunities for engaging with lived experiences remain limited due to practical constraints such as time, resources, and access to participants~\cite{baker2020systematic}. Prior work further suggests that accessibility compliance and automated evaluation do not necessarily capture how accessibility is experienced by users in practice~\cite{lewthwaite2016exploring, vinade2024educating, ismailova2022comparison}. Although automated tools are effective at identifying issues such as missing labels or contrast violations, they are less capable of capturing barriers related to navigation, comprehension, procedural complexity, and ongoing interaction processes~\cite{chakraborty2023artificial}. These limitations become especially significant in contexts where digital services are mandatory or near-mandatory, and users cannot easily opt out of digital participation.

\subsection{Older Adults, Cognitive Accessibility, and Digital Services}

As public and private services increasingly become digitally mediated, older adults are required to navigate digital systems for essential activities such as banking, healthcare, taxation, authentication, and communication~\cite{sakariassen2025multidimensional}. Prior research shows that older adults encounter barriers including navigation difficulties, uncertainty during multi-step procedures, memory-related challenges, and reduced confidence when systems change over time~\cite{kuzelewska2026elderly}. Cognitive accessibility is particularly important in this context, as users may struggle with comprehension, procedural complexity, inconsistent interfaces, and unfamiliar terminology~\cite{inal2025does}. Although accessibility standards address several of these issues, many barriers remain difficult to capture through compliance-oriented approaches because they emerge through interaction sequences and lived experiences rather than isolated interface violations~\cite{Inal,Inal2022}.

User-centered design traditions have long emphasized grounding design decisions in empirical understandings of users' experiences, needs, and contexts~\cite{sharpe2007interaction, w2013qualitative}. Design representations such as personas aim to bridge research and design by translating user experiences into actionable insights while maintaining a connection to lived realities~\cite{cooper2004inmates}. However, prior work also cautions that representations of users can oversimplify complex experiences and reduce contextual nuance~\cite{GlassCeiling,inal2025feel}. Similar concerns apply to accessibility evaluation practices that reduce accessibility to measurable checkpoints while overlooking how users experience digital systems across longer interaction processes~\cite{Inal}. Accessibility barriers in mandatory digital services are therefore not limited to interface problems, but also emerge as users navigate unfamiliar systems, recover from errors, interpret technical language, and adapt to continual system changes~\cite{Lindberg02102022}. Understanding accessibility in these contexts consequently requires moving beyond interface-level evaluation to examine how users experience digital services in everyday life.
\section{Method}

This paper reports on an exploratory survey aiming (1) to understand how older adults experience Norwegian web services that are legally required to be accessible, (2) to identify encountered barriers, and (3) to conceptualize these barriers as digital tensions emerging from misalignments between service design, system assumptions, and older adults’ capabilities. In this regard, a survey was designed and administered online over 4 weeks in March and April 2026. Data were collected from older adults aged 55 or older. The survey was distributed through relevant associations, communities, and social media. 

\subsection{Sample}

A total of 294 older adults participated in the study. Of the participants, 79 were male, and 215 were female. The age distribution of the participants ranged from 55-60 years (n~=~31), 61-70 years (n~=~77), 71-80 years (n~=~122), and 81-90 years (n~=~64). Concerning the educational levels of the participants, 8 had a primary school diploma, 32 had an upper secondary school degree, 46 had a vocational education, 133 had a Bachelor’s degree, 67 had a Master’s degree, and 8 had a PhD. On a 5-point scale from novice to expert, they rated
their level of knowledge concerning digital literacy skills to be 3.16 (SD~=~0.76).

\subsection{Data Analysis}
Survey questions, including older adults' demographics (age, gender, education, and perceived digital literacy skills), the web services they use, their strategies when learning to use a new digital service, and factors that prevent them from using web services, were analyzed descriptively. Thematic analysis ~\cite{braun2006using} was conducted on the open-ended question about barriers older adults encounter with web services. An inductive approach was followed, with tensions defined through the analysis process, without predefined categories at the outset. First, the answers were read multiple times to become familiar with the collected data. Initial codes and recurring patterns were documented, which would help define tensions and how they manifest as barriers. The codes were compared, organized, and grouped based on internal similarities. Then, relevant codes were systematized and classified into potential groups. In the final phase, emerging tensions were revised through multiple readings, improved, and named.

\begin{table*}[t]
\caption{Web services with level of frequency of use, representative
websites, and accessibility scores}
\label{tab:scores}
\centering

\begin{tabular}{
p{0.17\linewidth}  
p{0.10\linewidth}  
p{0.16\linewidth}  
p{0.30\linewidth}  
p{0.15\linewidth}  
}

\toprule

\textbf{Service \newline Categories}
&
\textbf{Usage Level}
&
\textbf{Usage Frequency} \newline Mean (SD)
&
\textbf{Representative Websites}
&
\textbf{Accessibility} \newline Mean (Min--Max)
\\

\midrule

BankID \newline Online Banking
&
\textbf{High}
&
4.17 (0.6) \newline 4.07 (0.7)
&
bankid, idporten, dnb, nordea, sparebank1, handelsbanken,
obos-banken, nordnet
&
97.9 (95--100)
\\

Mobile Payment \newline Streaming News \newline Health Services
&
\textbf{Moderate}
&
3.79 (0.81) \newline
3.51 (1.43) \newline
3.03 (0.79)
&
vipps, finn, vg, nrk, dagbladet, e24, tv2, sol, nettavisen,
aftenposten, helsenorge, aleris, volvat, dr.dropin, kry
&
97.4 (91--100)
\\

Online Shopping \newline
Tax Returns \newline
Welfare Benefits \newline
Video Consultation
&
\textbf{Low}
&
2.43 (0.91) \newline
2.31 (0.64) \newline
1.83 (0.81) \newline
1.38 (0.58)
&
tekk, elkjop, xxl, komplett, zalando, power, prisjakt, nav,
skatteateten, arbeidsplassen, altinn, ssb, udi, lanekassen,
oslo kommune, confrere, helserepons, eyr
&
92.7 (75--100)
\\

\bottomrule
\end{tabular}
\end{table*}

\section{Results}

We first report the use of web services, then the details of the four tensions identified in the free-text responses. 

\subsection{The Use of Web Services}

Participants rated their use of web services on a 5-point Likert scale from 1 (least) to 5 (most). Based on mean scores, we grouped the services into three levels of frequency of use, which were high, moderate, and low (see~\autoref{tab:scores}). BankID for access to public/private services and online banking were the most commonly used services, respectively, indicating a high frequency of use. These were followed by services for mobile payments and money transfers, streaming news, and health services such as prescriptions, test results, and appointment bookings, classified as moderately used services. The last group included low-frequency services, such as online shopping, tax return services, and relevant public services, digital services for welfare benefits and job support, and video consultations with a primary care physician or general practitioner.

We selected representative Norwegian websites associated with the identified service categories, representing above-mentioned web services. The list of websites was taken from top website ranking lists (e.g., SimilarWeb, Statista). We also included public-sector and private-sector digital services in areas such as health and finance. We then tested the accessibility status of these websites using Google Lighthouse ~\cite{google_lighthouse}, an online evaluation tool. The tool provides an accessibility score, based on WCAG, indicating the level of compliance. The average mean score for each group, along with the minimum and maximum scores, is given in~\autoref{tab:scores}.

When learning to use new digital services, the majority of the participants (n~=~213, 72\%) reported exploring on their own. This was followed by asking family (n~=~185, 63\%), friends (n~=~79, 27\%), and taking a relevant course (n~=~62, 21\%). Only 11 (4\%) participants reported not preferring to learn to use new digital services. Participants were also asked what prevents them from using the abovementioned web services more often. We grouped the responses into three key constraints on digital service use: individual, system-related, and contextual. Individual constraints consisted of fear of making mistakes (n~=~78, 27\%), lack of interest (n~=~60, 20\%), health challenges (e.g., vision, motor skills, memory) (n~=~20, 7\%), and lack of perceived benefit (n~=~13, 4\%). System-related constraints included concerns about privacy or data security (n~=~81, 28\%), perceived complexity of digital services (n~=~77, 26\%), and visual accessibility issues (e.g., small print or visual difficulties) (n~=~33, 11\%). Contextual constraints included lack of support or guidance (n~=~86, 29\%), preference for physical services (n~=~66, 22\%), and cost-related constraints (n~=~13, 4\%).

\subsection{Tensions}

This section outlines four main tensions, along with explanations of how each tension manifests. We also linked the tensions with relevant WCAG success criteria and elaborated on what needs to be done to meet the guidelines ~\cite{w3c2024wcag22} in the subsections (see~\autoref{tab:themes}). 

\begin{table*}[t]
\caption{Types of digital tension experienced by older adults in web services and relevant WCAG success criteria}
\label{tab:themes}
\centering

\begin{tabular}{@{}
  p{0.15\linewidth}
  p{0.30\linewidth}
  p{0.50\linewidth}
@{}}

\toprule
\textbf{Digital Tension}
&
\textbf{How the Tension Manifests}
&
\textbf{Relevant WCAG Success Criteria}
\\
\midrule

\textbf{Navigational}
&
Difficulty finding information, non-intuitive design, too many steps,
excessive complexity, interface changes, updates causing confusion.
&
1.4.3 Contrast (Minimum), 1.4.4 Resize Text, 1.4.10 Reflow,
2.2.4 Interruptions, 2.4.4 Link Purpose (In Context),
2.4.5 Multiple Ways, 2.4.6 Headings and Labels,
2.4.7 Focus Visible, 2.4.8 Location,
3.2.3 Consistent Navigation, 3.2.4 Consistent Identification,
3.3.2 Labels or Instructions
\\

\textbf{Semantic}
&
Difficulty understanding language, unclear instructions or explanations,
poor support services, memory-related challenges, and unhelpful error
messages.
&
2.2.6 Timeouts, 3.1.3 Unusual Words, 3.1.5 Reading Level,
3.2.6 Consistent Help, 3.3.1 Error Identification,
3.3.3 Error Suggestion,
3.3.4 Error Prevention (Legal, Financial, Data),
3.3.5 Help, 3.3.6 Error Prevention (All)
\\

\textbf{Procedural}
&
Password and PIN issues, authentication failures, login/access failure,
and multi-factor authentication complexity.
&
3.3.1 Error Identification, 3.3.3 Error Suggestion,
3.3.7 Redundant Entry,
3.3.8 Accessible Authentication (Minimum)
\\

\textbf{Temporal}
&
System downtime, service unavailability, slow or unstable performance,
device-related constraints (e.g., small screens, dexterity).
&
1.4.4 Resize Text, 1.4.10 Reflow,
2.2.1 Timing Adjustable, 2.2.2 Pause, Stop, Hide,
2.2.3 No Timing, 2.5.8 Target Size (Minimum),
3.2.1 On Focus, 3.2.2 On Input,
3.3.2 Labels or Instructions
\\

\bottomrule
\end{tabular}
\end{table*}

\subsubsection{Navigational Tension} The tension arises when the navigation structure is inconsistent across pages, overly complex to understand, or frequently updated, disrupting learned behaviour and requiring users to reestablish their proficiency when they return to use it after a period of inactivity. Participants reported feeling confused by the non-intuitive navigation. ``\textit{It's confusing. Can’t find my way}'' (P40). They metaphorically likened their experiences to being in a digital labyrinth. ``\textit{I’ve been led into labyrinths that I don’t understand} (P7)''. Accessibility guidelines aim to mitigate this by suggesting that there should be clear indicators given through multiple ways to help users understand where they are on the website (2.4.8 Location; 2.4.5 Multiple Ways). This could be provided by using a search function, site map, breadcrumbs, or menu items, so that users can engage with the content in a way that is most suitable for them, without getting lost. Further, to avoid confusion, all headings and labels in websites must be meaningful, descriptive, and consistent throughout the pages (2.4.6 Headings and Labels).

Participants were not happy with the design of web services. ``\textit{They call it `simpler', which it is NOT, or 'more modern', which often means gray text, less readability, and messy layouts} (P67)''. Regarding the use of color in web content, websites must follow contrast requirements for regular and large texts, as well as text in images (1.4.3 Contrast (Minimum)). However, the basic requirements may not guarantee optimal reading performance, meaning that websites should take responsibility for providing a ``truly'' accessible experience.

A lack of a clear signifier for the main design elements was one of the most common issues reported. ``\textit{Not all websites are self-explanatory… I struggle to find which buttons to press }(P38)''. Every field must have a descriptive and explicit label for clear guidance. Interactive elements should be reachable through keyboard navigation and include a visible focus indicator (2.4.7 Focus Visible) that stands out to help users identify the clickable elements on the page. There should also be instructions for complex processes (3.3.2 Labels or Instructions) and self-explanatory links that are embedded in the surrounding context (2.4.4 Link Purpose (In Context)) to scaffold users along their journey.  ``\textit{Often there is something in small print somewhere - or something you have to 'tick' to see other options - and it doesn't always come out well }(P54).'' 
 
Websites are required to be enlarged up to a certain percentage without distorting their content (1.4.4 Resize Text) or losing the functionality (1.4.10 Reflow). However, it was not the case for some web services. ``\textit{Sometimes I can't see the whole page, as I use enlarged font} (P57).'' Participants also reported being exposed to excessive information, making them feel overwhelmed: ``\textit{You drown in information you don’t need and get lost }(P141) and superfluous steps, causing complexity: ``\textit{Too many keystroke options… before I get to what I want to do }(P177).'' 

Frequently updated web content disrupted learned behavior. It is required not to interrupt users (2.2.4 Interruptions) with unnecessary updates that affect not only user performance but also assistive technologies that many users rely on to access digital services. Once participants had learned to use a web service, ``\textit{there was a change and [they] had to start over} (P143).'' Navigation elements such as menu items, links, etc., should remain consistent in location (3.2.3 Consistent Navigation) and identification (3.2.4 Consistent Identification) across pages after the updates, so that users can maintain their proficiency without spending excessive cognitive effort. The issue of updates being blocked was a major concern for several participants. ``\textit{The moment I learn to use a system, after a short time there is a change with the program, also starting over to adapt to new things, and that is the worst thing I experience} (P143).'' This made participants feel frustrated that their mastered skills were obsolete due to constant system changes, as when they saw an updated design, they had to reestablish their proficiency to be able to continue using the web service. ``\textit{Achieved skills are often outdated and require new skills as the service has changed since I last used it} (P164)''. 

\subsubsection{Semantic Tension} The tension emerges when there is a disconnection between the technical language in the web services and users' natural language, as well as the lack of system and human support when needed. Participants mostly reported difficulty in understanding the language used by web services. ``\textit{Concepts and jargon are used that I don't understand... Furthermore, everyone is expected to have a smartphone and be able to use all the apps} (P79)''. According to the guidelines, websites must provide content that is easy enough for users to understand (3.1.5 Reading Level) and technical, uncommon, or specialized terms must be explained in plain language (3.1.3 Unusual Words). 

Participants raised concerns that it might lead to more serious consequences when using highly critical web services, such as banking or healthcare, for transactions. The system should allow users to review and approve their actions before confirming them (3.3.4 Error Prevention (Legal, Financial, Data)), and language should not be the main barrier to completing these actions. Participants expected the web services to pay more attention to reducing the mismatch between their language and their mental models. Incomprehensible instructions, causing a lack of understanding, were described as a common issue as well. ``\textit{It seems that the help services often have a bit of difficulty with language. The commands are not always as logical to me} (P100).'' Due to the poor language use, error messages and system feedback were mostly unhelpful. ``\textit{I get error messages that I can’t use} (P21)''. It was not easy to understand even the problem itself. ``\textit{I don’t understand what I’m doing wrong} (P32)''. Websites should clearly inform users about the error by making the error field more salient (3.3.1 Error Identification) and by providing necessary information about solutions (3.3.3 Error Suggestion), so they can recover from it. 

One emerging barrier was a lack of technical support, as many participants complained about the inadequacy of digital-only support. They reported that chatbots ``\textit{don't understand the question} (P3)'' and there were ``\textit{too many responses... it can take an extremely long time} (P278)''. Older adults may need more time to achieve tasks. If there is a chance of losing data before they complete their actions, they should be informed about the details of the timeout process (2.2.6 Timeouts). Participants also described the importance of switching to human support whenever needed. ``\textit{Often you are referred to automated standard answers that do not provide answers to the questions you actually have, and that a living human could have answered} (P183).'' Providing help (3.3.5 Help) to users whenever they need through relevant pages, FAQ, instructions, contact options, etc., in a consistent and easily accessible way (3.2.6 Consistent Help) should be the priority of web services as digital services they provide seem to be the only option in several cases that older adults can not opt out.

\subsubsection{Procedural Tension} The tension emerges during login and verification processes in web services, particularly those involving BankID and password protocols. ``\textit{I can't get Bank ID to work at all. Error in my app? The bank should enter it} (P50).'' Participants reported the stress of security protocols and both physical and cognitive difficulty of logging in. ``\textit{Password is discarded even if it is correct} (P21)''. Websites should clearly inform users how to manage error messages by showing relevant input fields where the error occurred (3.3.1 Error Identification) and providing information on how to fix the error (3.3.3 Error Suggestion). 

Participants complained about multi-step verification, which added further complexity to user-system interaction, requiring them to switch between multiple windows during the login process. ``\textit{Prefers the chip… can’t find it back where I’m supposed to put the code} (P76)''. It is required to ease multi-step processes (3.3.7 Redundant Entry) for older adults by not asking them to provide the same information more than once when they are working on the same task, as they may have problems remembering what they have done before. 

Some participants mentioned age-related issues. ``\textit{Problems with remembering passwords, as I have to change them from time to time. The password should be secret, so I think it is unsafe to write it down} (P2).'' Websites should not rely on users' mental capabilities by forcing them to memorize and remember complex passwords or to solve difficult puzzles to log in to the system (3.3.8 Accessible Authentication (Minimum)). Instead, there should be easier options, such as biometric login or one-time codes that users can choose, as memory-related issues may become a real accessibility barrier, especially for older adults.

\subsubsection{Temporal Tension} The tension emerges when there are technical issues that mainly affect web services, such as service unavailability, system performance problems, or device-related constraints. Participants reported the instability of the web services they wanted to use. Sometimes the services ``\textit{can be down} (P15)” or ``\textit{work slowly} (P151).'' Even when a service was available, there were inconsistencies in the system performance. ``\textit{Logging in… lots of back and forth} (P126)''. 

There were some participants reporting difficulties with web services with quick changes between processes and screens, making the steps hard to follow: \textit{''The screen changes too fast, can't get my brain to follow that fast} (P244).'' Focus elements should not change content without user confirmation (3.2.1 On Focus, 3.2.2 On Input). As some users may need more time to complete a task (2.2.3 No Timing), the highest level of compliance is to avoid specific time limits. If some actions require time limits to complete, users should have control (2.2.1 Timing Adjustable) over those time limits by closing or extending them, depending on their preferences. Further, it is important to ensure that users have control over dynamic, auto-updating content (2.2.2 Pause, Stop, Hide), allowing them to focus on their tasks rather than being distracted by moving, scrolling, blinking, or updating design elements.

Device-related constraints, such as small screens when accessing a web service on mobile phones, and dexterity issues, were common among the participants. Problems with reading: ``\textit{due to age, it is difficult to find the right screen when I use bank ID to enter the password when using a mobile phone} (P266)'' or installation: ``\textit{difficult to install BANK-ID on the phone. Got help from my son} (P27)'' were also reported. As mobile devices have several challenges in terms of size, device capacity, interaction modalities, mobile versions of web services should be designed considering these limitations by especially providing enough size (2.5.8 Target Size (Minimum)) for interactive design elements as well as providing clear instructions for especially complex tasks or processes (3.3.2 Labels or Instructions), allow users to enlarge content and interface elements without losing information (1.4.4 Resize Text) and functionality (1.4.10 Reflow). This would especially help older adults gain independence by enabling them to use mobile versions of digital services without needing anyone's help.

\section{Discussion}

Our findings suggest that barriers experienced by older adults in digital services are not isolated usability and accessibility problems but systemic tensions arising from the mismatch between mandatory digital infrastructures and the realities of older adults’ everyday capabilities. In highly digitalized societies such as Norway, participation in essential activities, including banking, healthcare, taxation, and welfare, increasingly depends on the ability to use digital services~\cite{Parmiggiani2022,sakariassen2025multidimensional}. Although universal design of ICT is legally required across public and private sectors, accessibility in practice involves more than compliance with technical requirements. Age-related changes in memory, attention, vision, and dexterity affect how digital services are experienced and navigated in everyday contexts~\cite{w3c2025olderusers}. Our findings show that even services with high levels of accessibility compliance may remain difficult to use for older adults in practice.

The four tensions identified in our study reveal how digital services embed assumptions about users’ ability to continuously adapt to evolving systems. Navigational tensions emerged when participants were expected to relearn interfaces after updates, interpret inconsistent navigation structures, and maintain orientation across increasingly complex interfaces. Semantic tensions reflected assumptions that users share the technical language, terminology, and interpretive understanding embedded in digital systems. Procedural tensions emerged through authentication and multi-step verification processes that relied heavily on memory, procedural continuity, and error recovery. Temporal tensions highlighted how system speed, instability, time-sensitive interactions, and device-related constraints required users to continuously adapt to changing interaction conditions. Rather than emerging as isolated issues, these tensions collectively show how digital participation increasingly depends on users’ ability to continuously adapt to evolving digital systems.

Our findings further suggest a disconnect between compliance-oriented accessibility evaluation and lived accessibility in practice. Existing accessibility evaluation approaches primarily focus on detectable interface components such as labels, buttons, links, text alternatives, or contrast ratios. However, participants experienced barriers in interactional processes, including navigation, interpretation, authentication, maintaining proficiency after updates, and error recovery. In line with prior research~\cite{sayago2011ethnographical,kuzelewska2026elderly}, we observed that cognition- and comprehension-related challenges remain dominant barriers for older adults. Several barriers identified in our study are more closely related to WCAG AAA recommendations at the higher level, covering timing, terminology, process support, navigation, and error prevention. Yet these recommendations are not legally required for most web services. This generates a gap between minimum compliance requirements and the actual accessibility needs of older adults in everyday digital participation.

Our findings, therefore, highlight the importance of moving beyond checklist-based accessibility evaluation toward approaches that account for lived interaction over time. Accessibility in mandatory digital services is not only a property of interface components, but also of how systems support users in sustaining understanding, navigation, procedural competence, and confidence across repeated interactions. We suggest that accessibility evaluation should place greater emphasis on actual user performance and cognitive accessibility in real-world contexts, particularly for services that citizens cannot meaningfully opt out of using. Supporting older adults' participation in digital society, therefore, requires accessibility approaches that consider how users interpret, navigate, and adapt to evolving digital systems, rather than focusing solely on compliance.
\section{Conclusion}

This paper examined the tensions resulting from systemic misalignments between service design, system assumptions, and older adults’ capabilities. These tensions, namely navigational, semantic, procedural, and temporal, describe how accessibility barriers emerge throughout ongoing interaction with digital services. We pointed out that although public and private sector services are legally required to comply with the universal design requirements of Norwegian ICT regulations, meeting technical requirements does not necessarily reflect the real-life accessibility experiences of older adults. Therefore, designers, developers, service providers, and policymakers should not rely solely on checklist-based accessibility compliance to help older adults engage with digital services independently. More broadly, particularly in highly digitalized countries with an increasingly ageing population, such as Norway, our work highlights the need to rethink how to move beyond minimum compliance in digital services provided by public and private sector bodies, by addressing cognitive accessibility challenges related to attention, understanding, perception, and memory on the web. Our analysis is based on responses from 294 older adults, limiting the diversity of contexts and experiences. The findings of our study should be considered as indicative rather than representative. Future research should expand its scope to include digital platforms, target users, and contexts to better understand real-life tensions between what service providers offer and what users demand.


%


\bibliographystyle{ACM-Reference-Format}
\bibliography{references}

@article{GlassCeiling,
title = {“I’ve never seen a glass ceiling better represented”: Bias and gendering in LLM-generated synthetic personas from a participatory design perspective},
journal = {International Journal of Human-Computer Studies},
volume = {205},
pages = {103651},
year = {2025},
issn = {1071-5819},
doi = {https://doi.org/10.1016/j.ijhcs.2025.103651},
url = {https://www.sciencedirect.com/science/article/pii/S1071581925002083},
author = {Helena A. Haxvig and Vincenzo D’Andrea and Maurizio Teli}
}

@article{ismailova2022comparison,
  title={Comparison of online accessibility evaluation tools: an analysis of tool effectiveness},
  author={Ismailova, Rita and Inal, Yavuz},
  journal={IEEe Access},
  volume={10},
  pages={58233--58239},
  year={2022},
  publisher={IEEE}
}

@software{google_lighthouse,
  author       = {{Google}},
  title        = {Lighthouse},
  year         = {2026},
  url          = {https://developer.chrome.com/docs/lighthouse},
  note         = {Accessed: 2026-05-13}
}

@inproceedings{Inal,
author = {Inal, Yavuz and Guribye, Frode and Rajanen, Dorina and Rajanen, Mikko and Rost, Mattias},
title = {Perspectives and Practices of Digital Accessibility: A Survey of User Experience Professionals in Nordic Countries},
year = {2020},
isbn = {9781450375795},
publisher = {Association for Computing Machinery},
address = {New York, NY, USA},
url = {https://doi.org/10.1145/3419249.3420119},
doi = {10.1145/3419249.3420119},
booktitle = {Proceedings of the 11th Nordic Conference on Human-Computer Interaction: Shaping Experiences, Shaping Society},
articleno = {63},
numpages = {11},
location = {Tallinn, Estonia},
series = {NordiCHI '20}
}

@article{chakraborty2023artificial,
title = {Artificial Intelligence: The road ahead for the accessibility of persons with Disability},
journal = {Materials Today: Proceedings},
volume = {80},
pages = {3757-3761},
year = {2023},
note = {SI:5 NANO 2021},
issn = {2214-7853},
doi = {https://doi.org/10.1016/j.matpr.2021.07.374},
url = {https://www.sciencedirect.com/science/article/pii/S2214785321052330},
author = {Nilanjan Chakraborty and Yogesh Mishra and Ripon Bhattacharya and Bhupal Bhattacharya}
}

@inproceedings{antonelli2018survey,
author = {Antonelli, Humberto Lidio and Rodrigues, Sandra Souza and Watanabe, Willian Massami and de Mattos Fortes, Renata Pontin},
title = {A survey on accessibility awareness of Brazilian web developers},
year = {2018},
isbn = {9781450364676},
publisher = {Association for Computing Machinery},
address = {New York, NY, USA},
url = {https://doi.org/10.1145/3218585.3218598},
doi = {10.1145/3218585.3218598},
booktitle = {Proceedings of the 8th International Conference on Software Development and Technologies for Enhancing Accessibility and Fighting Info-Exclusion},
pages = {71–79},
numpages = {9},
location = {Thessaloniki, Greece},
series = {DSAI '18}
}

@article{farrelly2011practitioner,
author = {Glen Farrelly},
title ={Practitioner barriers to diffusion and implementation of web accessibility},

journal = {Technology and Disability},
volume = {23},
number = {4},
pages = {223-232},
year = {2011},
doi = {10.3233/TAD-2011-0329},

URL = { 
 https://journals.sagepub.com/doi/abs/10.3233/TAD-2011-0329
}
}

@inproceedings{soares2020we,
author = {Soares Guedes, Leandro and Landoni, Monica},
title = {How Are We Teaching and Dealing with Accessibility? A Survey From Switzerland},
year = {2021},
isbn = {9781450389372},
publisher = {Association for Computing Machinery},
address = {New York, NY, USA},
url = {https://doi.org/10.1145/3439231.3440610},
doi = {10.1145/3439231.3440610},
booktitle = {Proceedings of the 9th International Conference on Software Development and Technologies for Enhancing Accessibility and Fighting Info-Exclusion},
pages = {141–146},
numpages = {6},
location = {Online, Portugal},
series = {DSAI '20}
}

@article{inal2018computer,
  title={How do computer engineering students construe usability and accessibility? A comparative study between Turkey and Kyrgyzstan},
  author={Inal, Yavuz and Ismailova, Rita},
  journal={Tehni{\v{c}}ki vjesnik},
  volume={25},
  number={5},
  pages={1339--1347},
  year={2018},
  publisher={Sveu{\v{c}}ili{\v{s}}te u Slavonskom Brodu, Stojarski fakultet},
  doi = {10.17559/TV-20170205083820},

URL = { 
 https://doi.org/10.17559/TV-20170205083820
},
}

@article{inal2019web,
  title={Web accessibility in Turkey: awareness, understanding and practices of user experience professionals},
  author={Inal, Yavuz and R{\i}zvano{\u{g}}lu, Kerem and Yesilada, Yeliz},
  journal={Universal Access in the Information Society},
  volume={18},
  number={2},
  pages={387--398},
  year={2019},
  publisher={Springer},
    url = {https://doi.org/10.1007/s10209-017-0603-3},
doi = {10.1007/s10209-017-0603-3},
}

@inproceedings{katsanos2009web,
  title={Web accessibility: Design of an educational system to support guidelines learning},
  author={Katsanos, Christos and Tsakoumis, Athanasios and Avouris, Nikolaos},
  booktitle={Proc. of the 13th Pan-Hellenic Conference on Informatics (PCI)},
  pages={155--164},
  year={2009}
}

@article{putnam2016best,
author = {Putnam, Cynthia and Dahman, Maria and Rose, Emma and Cheng, Jinghui and Bradford, Glenn},
title = {Best Practices for Teaching Accessibility in University Classrooms: Cultivating Awareness, Understanding, and Appreciation for Diverse Users},
year = {2016},
issue_date = {May 2016},
publisher = {Association for Computing Machinery},
address = {New York, NY, USA},
volume = {8},
number = {4},
issn = {1936-7228},
url = {https://doi.org/10.1145/2831424},
doi = {10.1145/2831424},
journal = {ACM Trans. Access. Comput.},
month = mar,
articleno = {13},
numpages = {26}
}

@article{inal2025feel,
  title={I feel for you! The effect of empathy exercises on accessibility awareness},
  author={Inal, Yavuz},
  journal={Universal Access in the Information Society},
  volume={24},
  number={1},
  pages={535--542},
  year={2025},
  publisher={Springer},
  url = {https://doi.org/10.1007/s10209-023-01081-z},
doi = {10.1007/s10209-023-01081-z},
}

@article{braun2006using,
author = {Virginia Braun and Victoria Clarke},
title = {Using thematic analysis in psychology},
journal = {Qualitative Research in Psychology},
volume = {3},
number = {2},
pages = {77--101},
year = {2006},
publisher = {Routledge},
doi = {10.1191/1478088706qp063oa},
URL = { 
    
        https://doi.org/10.1191/1478088706qp063oa
},
}

@inproceedings{lewthwaite2016exploring,
author = {Lewthwaite, Sarah and Sloan, David},
title = {Exploring pedagogical culture for accessibility education in computing science},
year = {2016},
isbn = {9781450341387},
publisher = {Association for Computing Machinery},
address = {New York, NY, USA},
url = {https://doi.org/10.1145/2899475.2899490},
doi = {10.1145/2899475.2899490},
booktitle = {Proceedings of the 13th International Web for All Conference},
articleno = {3},
numpages = {4},
location = {Montreal, Canada},
series = {W4A '16}
}

@inproceedings{ludi2018teaching,
author = {Ludi, Stephanie and Huenerfauth, Matt and Hanson, Vicki and Rajendra Palan, Nidhi and Conn, Paula},
title = {Teaching Inclusive Thinking to Undergraduate Students in Computing Programs},
year = {2018},
isbn = {9781450351034},
publisher = {Association for Computing Machinery},
address = {New York, NY, USA},
url = {https://doi.org/10.1145/3159450.3159512},
doi = {10.1145/3159450.3159512},
booktitle = {Proceedings of the 49th ACM Technical Symposium on Computer Science Education},
pages = {717–722},
numpages = {6},
location = {Baltimore, Maryland, USA},
series = {SIGCSE '18}
}

@inproceedings{vinade2024educating,
author = {Vinad\'{e}, Renata and Marczak, Sabrina},
title = {Educating for Accessibility: Insights into Knowledge Gaps and Practical Challenges in Software Engineering},
year = {2024},
isbn = {9798400717772},
publisher = {Association for Computing Machinery},
address = {New York, NY, USA},
url = {https://doi.org/10.1145/3701625.3701689},
doi = {10.1145/3701625.3701689},
booktitle = {Proceedings of the XXIII Brazilian Symposium on Software Quality},
pages = {646–656},
numpages = {11},
location = {
},
series = {SBQS '24}
}

@inproceedings{baker2020systematic,
author = {Baker, Catherine M. and El-Glaly, Yasmine N. and Shinohara, Kristen},
title = {A Systematic Analysis of Accessibility in Computing Education Research},
year = {2020},
isbn = {9781450367936},
publisher = {Association for Computing Machinery},
address = {New York, NY, USA},
url = {https://doi.org/10.1145/3328778.3366843},
doi = {10.1145/3328778.3366843},
booktitle = {Proceedings of the 51st ACM Technical Symposium on Computer Science Education},
pages = {107–113},
numpages = {7},
location = {Portland, OR, USA},
series = {SIGCSE '20}
}

@inproceedings{el2020presenting,
author = {El-Glaly, Yasmine and Shi, Weishi and Malachowsky, Samuel and Yu, Qi and Krutz, Daniel E.},
title = {Presenting and evaluating the impact of experiential learning in computing accessibility education},
year = {2020},
isbn = {9781450371247},
publisher = {Association for Computing Machinery},
address = {New York, NY, USA},
url = {https://doi.org/10.1145/3377814.3381710},
doi = {10.1145/3377814.3381710},
booktitle = {Proceedings of the ACM/IEEE 42nd International Conference on Software Engineering: Software Engineering Education and Training},
pages = {49–60},
numpages = {12},
location = {Seoul, South Korea},
series = {ICSE-SEET '20}
}

@book{cooper2004inmates,
  title={The inmates are running the asylum: Why high-tech products drive us crazy and how to restore the sanity},
  author={Cooper, Alan and others},
  volume={2},
  year={2004},
  publisher={Sams Indianapolis}
}

@inproceedings{w2013qualitative,
  title={Qualitative inquiry and research design choosing among five approaches},
  author={W John, Creswell},
  year={2013},
  organization={Library of Congress Cataloging-in-Publication Data}
}

@article{sharpe2007interaction,
  title={Interaction design: beyond human-computer interaction 2nd ed},
  author={Sharpe, H and Rogers, Y and Preece, J},
  journal={H Sharpe, Y Rogers, and J Preece},
  year={2007}
}

@inproceedings{Inal2022,
author = {Inal, Yavuz and Mishra, Deepti and Torkildsby, Anne Britt},
title = {An Analysis of Web Content Accessibility of Municipality Websites for People with Disabilities in Norway: Web Accessibility of Norwegian Municipality Websites},
year = {2022},
isbn = {9781450396998},
publisher = {Association for Computing Machinery},
address = {New York, NY, USA},
url = {https://doi.org/10.1145/3546155.3547272},
doi = {10.1145/3546155.3547272},
booktitle = {Nordic Human-Computer Interaction Conference},
articleno = {65},
numpages = {12},
location = {Aarhus, Denmark},
series = {NordiCHI '22}
}

@article{sakariassen2025multidimensional,
  title={Multidimensional digital vulnerability among older adults},
  author={Sakariassen, Hilde},
  journal={Nordicom Review},
  volume={46},
  number={2},
  pages={137--160},
  year={2025},
  publisher={De Gruyter Poland}
}

@article{inal2025does,
  title={Does the law make a difference? a longitudinal study on accessibility compliance of Norwegian municipality websites},
  author={Inal, Yavuz and Torkildsby, Anne Britt},
  journal={Universal Access in the Information Society},
  volume={24},
  number={2},
  pages={1845--1855},
  year={2025},
  publisher={Springer}
}

@article{kuzelewska2026elderly,
  title={The elderly digital divide: Digital exclusion versus the right not to use the internet},
  author={Ku{\.z}elewska, El{\.z}bieta and Tomaszuk, Mariusz and Malinowski, Damian},
  journal={International Journal for the Semiotics of Law-Revue internationale de S{\'e}miotique juridique},
  volume={39},
  number={4},
  pages={1365--1384},
  year={2026},
  publisher={Springer}
}

@article{Lindberg02102022,
author = {Jens Lindberg and Elin Kvist and Simon Lindgren},
title = {The Ongoing and Collective Character of Digital Care for Older People: Moving Beyond Techno-Determinism in Government Policy},
journal = {Journal of Technology in Human Services},
volume = {40},
number = {4},
pages = {357--378},
year = {2022},
publisher = {Routledge},
doi = {10.1080/15228835.2022.2144588},


URL = { 
    
        https://doi.org/10.1080/15228835.2022.2144588
    
    

},
eprint = { 
    
        https://doi.org/10.1080/15228835.2022.2144588
    
    

}



}

@misc{w3c2025olderusers,
  author       = {{W3C Web Accessibility Initiative}},
  title        = {Older Users and Web Accessibility: Meeting the Needs of Ageing Web Users},
  year         = {2025},
  howpublished = {\url{https://www.w3.org/WAI/older-users/}},
  note         = {World Wide Web Consortium. Accessed: 2026-05-11}
}

@inproceedings{petrie2007relationship,
  title={The relationship between accessibility and usability of websites},
  author={Petrie, Helen and Kheir, Omar},
  booktitle={Proceedings of the SIGCHI conference on Human factors in computing systems},
  pages={397--406},
  year={2007}
}

@article{sayago2011ethnographical,
      title={An ethnographical study of the accessibility barriers in the everyday interactions of older people with the web},
      author={Sayago, Sergio and Blat, Josep},
      journal={Universal Access in the Information Society},
      volume={10},
      number={4},
      pages={359--371},
      year={2011},
      publisher={Springer}
}

@misc{uutilsynet2025wcag,
    author = {{Norwegian Digitalisation Agency}},
    title = {Guidance and information},
    year = {2025},
    howpublished = {\url{https://www.uutilsynet.no/english/guidance-and-information/905}},
    note = {Accessed: 2026-05-11}
}

@misc{w3c2024wcag22,
    author = {{World Wide Web Consortium}},
    title = {Web Content Accessibility Guidelines (WCAG) 2.2},
    year = {2024},
    howpublished = {\url{https://www.w3.org/TR/WCAG22/}},
    note = {W3C Recommendation. Accessed: 2026-05-11}
}

@Inbook{Parmiggiani2022,
author="Parmiggiani, Elena
and Mikalef, Patrick",
editor="Mikalef, Patrick
and Parmiggiani, Elena",
title="The Case of Norway and Digital Transformation over the Years",
bookTitle="Digital Transformation in Norwegian Enterprises                                ",
year="2022",
publisher="Springer International Publishing",
address="Cham",
pages="11--18",
isbn="978-3-031-05276-7",
doi="10.1007/978-3-031-05276-7_2",
url="https://doi.org/10.1007/978-3-031-05276-7_2"
}

\end{document}